\documentclass[
reprint,
showpacs,
nofootinbib,
superscriptaddress,
longbibliography,
amsmath,amssymb,prd]{revtex4-1}

\usepackage{graphicx}% include figure files
\usepackage{dcolumn}% align table columns on decimal point
\usepackage{siunitx}
\usepackage{bm}% bold math
\usepackage{esdiff}% for derivatives
\usepackage{hyperref}% add hypertext capabilities
\usepackage[mathlines]{lineno}% enable numbering of text and display math
\usepackage{subfigure}
\begin{document}
\title{The Relativistic Geoid: Gravity Potential and Relativistic Effects}

\author{Dennis Philipp}
\affiliation{ZARM, University of Bremen, 28359 Bremen, Germany}
\affiliation{Fraunhofer MEVIS, 28359 Bremen, Germany}
\author{Eva Hackmann}
\affiliation{ZARM, University of Bremen, 28359 Bremen, Germany}
\author{Claus L\"ammerzahl}
\affiliation{ZARM, University of Bremen, 28359 Bremen, Germany}
\affiliation{DLR-Institute for Satellite Geodesy and Inertial Sensing,
c/o University of Bremen, Am Fallturm 2, 28359 Bremen, Germany}
\affiliation{Institute of Physics, Carl von Ossietzky University Oldenburg, 26111 Oldenburg, Germany}
\author{J{\"u}rgen M{\"u}ller}
\affiliation{IfE, University of Hannover, 30167 Hannover, Germany}
\affiliation{DLR-Institute for Satellite Geodesy and Inertial Sensing,
c/o Leibniz Universität Hannover, Welfengarten 1, 30167 Hannover, Germany}

%%
%-- abstract
%%

\begin{abstract} 
The Earth's geoid is one of the most essential and fundamental concepts to provide a gravity field-related height reference in geodesy and associated sciences. 
To keep up with the ever-increasing experimental capabilities and to consistently interpret high-precision measurements without any doubt, a relativistic treatment of geodetic notions (including the geoid) within Einstein's theory of General Relativity is inevitable. 
Building on the theoretical construction of isochronometric surfaces and the so-called redshift potential for clock comparison, we define a relativistic gravity potential as a generalization of (post-)Newtonian notions. 
This potential exists in any stationary configuration with rigidly co-rotating observers, and it is the same as realized by local plumb lines. 
In a second step, we employ the gravity potential to define the relativistic geoid in direct analogy to the Newtonian understanding. In the respective limit, the framework allows to recover well-known (post-) Newtonian results. 
For a better illustration and proper interpretation of the general relativistic gravity potential and geoid, some particular examples are considered. 
Explicit results are derived for exact vacuum solutions to Einstein's field equation as well as a parametrized post-Newtonian model. 
Comparing the Earth's Newtonian geoid to its relativistic generalization is a very subtle problem, but of high interest.
An isometric embedding into Euclidean three-dimensional space is an appropriate solution and allows a genuinely intrinsic comparison. With this method, the leading-order differences are determined, which are at the mm-level. 
\end{abstract} 

%pacs numbers
%keywords
\maketitle

%\tableofcontents

%%
%-- introduction
%%

\section{\label{sec:introduction} Introduction}
The ever-increasing technological capabilities allow us to perform gravity and clock measurements with incredible accuracy.
On the one hand, high-precision satellite missions such as GRACE/GRACE-FO allow to deduce properties of the Earth's gravity field, its changes on various time scales, and the investigation of underlying phenomena \cite{Tapley:2004, Tapley:2019, Abich:2019}.
On the other hand, Earth-bound clock comparison networks or portable optical atomic clocks are used in the framework of chronometric geodesy \cite{Takano:2016, Mueller:2017, Kopeikin:2016b, Delva:2018}. 
One of the central notions that is to be determined by such geodetic measurements is the Earth's geoid -- its mathematical figure as the German mathematician C.F.\ Gauss has termed it.
Geoid determination with high accuracy is necessary for, e.g., national and global height systems and is related to various applications such as GNSS.

To thoroughly explain the outcome of contemporary and future geodetic missions at the cutting edge of available accuracy, we have to keep up at the theoretical level.
Consequently, geodetic notions and concepts must be developed within a relativistic theory of gravity.
The best available framework, consistent with all tests, is Einstein's theory \cite{Will:2014}. Therefore, it is our goal to generalize known geodetic concepts and define all notions intrinsically within General Relativity.
To consistently interpret high-precision measurements without any doubt, a relativistic derivation of geodetic notions within General Relativity and beyond post-Newtonian gravity is inevitable. This is true in particular for the conceptual formulation of relativistic geodesy and model building in a top-down approach.

In this work, we define a relativistic gravity potential, which generalizes the Newtonian one. It exists for any stationary configuration\footnote{By configuration we mean a spacetime model for the Earth's exterior.} with observers on isometric congruences, i.e.\ observers who rigidly co-rotate with the Earth.
Its definition is based on the philosophy of Bjerhammar \cite{Bjerhammar:1985,Bjerhammar:1986} and Soffel \emph{et al.}\ \cite{Soffel:1988} as well as results on the time-independent redshift potential in Ref.\  \cite{Philipp:2017}.
It allows geodetic notions such as the geoid to be defined and generalized in an intrinsic general relativistic manner and with well-defined weak-field limits.
Moreover, it can be used to calculate the outcome of redshift and acceleration measurements, and it is realized by clock comparison as well as the determination of local plumb lines.

One significant result is the direct comparison of the conventional Newtonian geoid and its relativistic generalization.
Such a comparison is of relevance for geodesy, but involves a lot of subtle points. 
In particular, there is some gauge freedom in the choice of constants, different applicable conventions, and the more crucial geometrical problem of comparing objects that live in different geometries. We show how to perform this comparison using an isometric embedding and different conventions.

The structure of this work is as follows. We start with a short recapitulation of the conventional geoid and results in the literature in Sec.\ \ref{sec:prevResults}.
After introducing the relativistic gravity potential in Sec.\ \ref{sec:relGravityPotential}, it is employed to give a definition of the general relativistic geoid in direct analogy to the Newtonian case. We show how the potential can be used to express clock comparison as well as acceleration measurements between observers on the Earth's surface.
Our definition of the geoid is such that it is the surface which is locally orthogonal to plumb lines and coincides with the surface of vanishing mutual redshifts of standard clocks on Killing congruences.
Therefore, different measurements can contribute to its realization in data fusion.

In Sec.\ \ref{Sec:Examples} the definitions are applied to some particular spacetime examples for illustration of the concepts and proper interpretation. 
A first-order parametrized post-Newtonian metric is used to show that results in the literature are embedded into the present framework.
Exact expressions for the Schwarzschild spacetime, the quadrupolar Erez-Rosen spacetime, general asymptotically flat Weyl metrics, and the Kerr spacetime are derived as well. 
Thus, the effects of the relativistic monopole, the quadrupole, and higher-order multipoles in axisymmetric configurations can be analyzed order by order.
Moreover, approximating the Earth's exterior spacetime by a suitably constructed Kerr metric allows us to consider gravitomagnetic contributions.

In the last part, we compare the conventional Newtonian geoid to its relativistic generalization.
Involved problems and subtleties are overcome by an isometric embedding of the relativistic geoid surface into Euclidean three-dimensional space, and we show that the leading-order difference, for a suitable convention, is about $2\,$mm due to the relativistic monopole.
This embedding is not only an academic endeavor but necessary to overcome coordinate-dependent effects.

%%
%----------------------------------------------------------------------
%%

\section{\label{sec:prevResults} Conventional Understanding and Previous Results}
In this section, the conventional understanding of the geoid in Newtonian gravity as well as generalizations that exist so far within (approximate) relativistic frameworks are briefly summarized. 
We start with the Newtonian geoid and consider the post-Newtonian extension thereafter.
In addition, we summarize helpful references in geodetic and general relativistic literature. 
\subsection{Newtonian Geoid}
In Newtonian gravity, the geoid is defined as one particular level surface of the gravity potential\footnote{Note that in conventional geodesy, the gravitational potential is usually denoted by $V$, whereas the centrifugal potential is denoted by $Z$.} \cite{TorgeMueller:2012},
\begin{align}
	W\big(\vec{X}\big) := U\big(\vec{X}\big) + V\big(\vec{X}\big) \, ,
\end{align}
where $U(\vec{X})$ is the Newtonian gravitational potential and $V(\vec{X})$ is the centrifugal potential experienced by rigidly co-rotating observers on the Earth's surface.
We deliberately make the distinction between gravitation and gravity here to match geodetic notions and conventions.
In Earth-centered global spherical coordinates $(R,\Theta,\Phi)$, we then have for the centrifugal potential,
\begin{align}
	V\big( \vec{X} \big) := - \dfrac{1}{2} \omega^2 d_z^2 = - \dfrac{1}{2} \omega^2 R^2 \sin^2 \Theta \, ,
\end{align} 
and the expansion of the gravitational potential into spherical harmonics reads
\begin{multline}
	\label{Eq:NewtonianPotentialExpansion}
	U(R,\Theta,\Phi) = - \dfrac{GM}{R} \sum_{l=0}^\infty \sum_{m=0}^l \left( \dfrac{R_{\text{ref}}}{R} \right)^l P_{lm} (\cos \Theta) \\
	\times 
	\left[ C_{lm} \cos (m \Phi) + S_{lm} \sin (m \Phi) \right] \, .
\end{multline}
In the equations above, $\omega$ is the angular frequency of the Earth's rotation, $R_{\text{ref}}$ is some chosen reference radius, and $d_z$ is the distance to the rotation axis which points into the $z$-direction.
The $P_{lm}$ are the Legendre functions of degree $l$ and order $m$, and $C_{lm}, S_{lm}$ are multipole expansion coefficients.
The gravitational potential is a solution of Poisson's equation (Laplace's equation outside the sources) and adapted to the condition that $U \to 0$ for $R \to \infty$.
Note that in our sign convention the gravity potential is always negative since it refers to an attractive force.
Under the assumption of axial symmetry, the expansion  simplifies to
\begin{align}
	\label{Eq:NewtonianPotentialExpansionAxisym}
	U(R,\Phi) = - \dfrac{GM}{R} \sum_{l=0}^\infty J_l \left( \dfrac{R_{\text{ref}}}{R} \right)^l P_l (\cos \Theta)\, ,
\end{align}
where the $J_l$ are axially symmetric multipole moments.
A suitable definition of the Newtonian geoid now is the following.

\emph{Definition:} The Earth's geoid is defined by the level surface of the gravity potential ${W \big( \vec{X} \big)}$ such that
\begin{align}
	\label{Eq:geoidDefNewton}
	- W \big( \vec{X} \big) \big|_{\text{geoid}} = W_0 = \text{constant} \, ,
\end{align}
with a constant $W_0 = \SI{6.26368534e7}{\square\meter \per \square\second}$, which complies with modern conventions \cite{TorgeMueller:2012, Sanchez:2016}.
The numerical value of $W_0$ is chosen in a way to have the geoid coincide best with the mean sea level at rest and a history of measurements and previous conventions, see Ref.\ \cite{Sanchez:2016}.
In contrast to usual geodetic formulations, we use a negative potential such that our convention in Eq.\ \eqref{Eq:geoidDefNewton} differs by a sign. However, we want to keep the numerical value of $W_0$
 to be strictly positive.
 
\subsection{Post-Newtonian and General Relativistic Approaches}
The first attempt to define a relativistic geoid was undertaken by Bjerhammar \cite{Bjerhammar:1985, Bjerhammar:1986} in 1985. He defined the geoid to be the surface on which ``precise clocks run with the same speed'', but most of the considerations involve approximations of order $c^{-2}$ and special relativistic results.
Inspired by Bjerhammar's approach, which, however, lacks some formal and mathematical clarity, we give a rigorous general relativistic definition of a gravity potential and the geoid, based on clock comparison without approximations. Thus, we go beyond Bjerhammar's considerations and generalize his ideas. 
The essential steps to do so have already been outlined in Refs.\ \cite{Philipp:2017, Philipp:2017b, Philipp:2019}, in which the relativistic geoid is defined in terms of isochronometric surfaces, the level sets of a so-called stationary redshift potential for Killing congruences. Furthermore, the analysis of timelike isometric congruences as the worldlines of observers, or equivalently the Earth's matter constituents, allows to derive an acceleration potential. It coincides with the redshift potential and contributes yet another possibility of how to determine the relativistic geoid as the surface orthogonal to the direction of local plumb lines.
We will, to a large extent, use the results in the references above and incorporate them into the definition of a general relativistic gravity potential in the next section.

Several authors also investigated the Earth's geoid in a post-Newtonian framework; see, e.g., Refs.\ \cite{Soffel:1988, Kopeikin:1991,Kopeikin:2007,Kopeikin:2015}.
The principle idea is to use potentials defined at the order of $1/c^2$ in a post-Newtonian approximation of General Relativity. 
The relativistic geoid can then be defined by a special level surface again but is valid to first order only.
In this context, the notions of the u-geoid and a-geoid appear, related to definitions in terms of clocks and their comparison on the one hand and accelerations of observers on the other hand, see Ref.\ \cite{Soffel:1988}. They are found to coincide at the first order in post-Newtonian gravity.
However, the choice of a particular surface w.r.t.\ the Newtonian limit as well as an intrinsic relativistic understanding is generally not elaborated. In the present formalism, it can be proven that the u- and a-geoids generically coincide in General Relativity without any approximation involved; see Refs. \cite{Philipp:2017, Philipp:2019}.

Concerning a general relativistic treatment, the work in Ref.\ \cite{Oltean:2015} must be mentioned, in which the authors define the geoid in terms of quasi-local frames.
Furthermore, in Ref.\ \cite{Kopeikin:2015}, an exact definition of the geoid and a generalization of the disturbing potential in General Relativity are analyzed via perturbations of a chosen background manifold. 

For the geodetic community also the difference between Newtonian and generalized notions is of great interest but a thorough comparison is not included in the available literature.
As we will show in the next sections, in such a comparison some subtle details are involved, but they can be handled genuinely within a general relativistic framework and by differential geometric methods without approximations. 
Thus, intrinsic geometric properties are conserved and not spoiled by an approximative description.

%%
%----------------------------------------------------------------------
%%

\section{\label{sec:relGravityPotential} Relativistic Gravity Potential and the geoid}

To arrive at an intrinsic relativistic understanding of a stationary object's geoid, we build on some of the ideas in Refs.\ \cite{Soffel:1988, Kopeikin:2015} together with our previous results in Ref.\ \cite{Philipp:2017}.
Thereupon, we present a framework that is consistent within General Relativity without any approximation and allows previously known (post-) Newtonian results to be recovered in the respective limits.

The next section contains a summary of significant results regarding clock comparison, isometric observer congruences, and isochronometric surfaces.
Thereupon, we define a relativistic gravity potential that is, in turn, used to define the relativistic geoid in analogy to the Newtonian understanding.

We use SI units to explicitly see how the speed of light $c$ and Newton's gravitational constant $G$ enter into the formulae. 
Greek indices are spacetime indices and shall range from 0 to 3, whereas Latin indices are purely spatial indices and take values from 1 to 3.

\subsection{Observer Congruence and Redshift Potential} 
Any stationary metric can be written in the form
\begin{align}
	\label{Eq:metricStationary}
	g = e^{2 \phi (x) } \left[ -(c \, \mathrm{d}t + \alpha_a(x) \mathrm{d}x^a)^2 + \alpha_{ab}(x) \mathrm{d}x^a \mathrm{d}x^b \right] \, ,
\end{align}
in which the coordinates ${(x^0=c\,t,x^i)}$ for $i=1:3$ are assumed to be co-rotating and are adapted to the symmetry. 
Let us assume that Eq.\ \eqref{Eq:metricStationary} describes the exterior of some astronomical object of interest. Given the geodesy of our planet, we use this metric to build a model of the Earth's exterior spacetime.
Because of the lack of better wording, we will also speak of the geodesy of other objects and define the geoid in a way that works generally.
Now, we think of observers that rigidly co-rotate with the object of interest.\footnote{Alternatively, also the constituents of the Earth in a rigid model follow the same worldlines.} 
They transport standard clocks along their worldlines which are parametrized by proper time, respectively.
Since the coordinates are co-rotating, these observers are hovering on fixed positions (${\mathrm{d}x^i = 0}$) or might be attached to the surface if there is any.
Such observers are described by integral curves of the Killing vector field $\xi_{(1)} := \partial_t$.
For the Earth, this means we consider in particular observers (measurement stations equipped with atomic clocks and gravimeters) attached to its physical surface.
The worldlines of these observers form an isometric (Killing) congruence, and there are some important conclusions.
As shown in Ref.\ \cite{Philipp:2017}, for a spacetime with a metric in the form of Eq.\ \eqref{Eq:metricStationary} there exists a time-independent redshift potential. It describes the relative frequency difference between any two standard clocks on worldlines in the congruence.
This redshift potential is given by the scalar function $\phi(x)$ and two observers on worldlines $\gamma_1$ and $\gamma_2$, respectively, determine their mutual redshift
\begin{align}
	\label{Eq:redshiftPhi}
	1 + z := \dfrac{\nu_1}{\nu_2} = \exp(\phi |_{\gamma_2} - \phi |_{\gamma_1}) =: \exp(\Delta \phi) \, ,
\end{align}
in which $\nu_{1,2}$ is the frequency of an exchanged light signal as seen by the respective observer. All observer worldlines in the congruence are described by the tangent vector field $u = \exp(-\phi) \xi_{(1)}$.
The redshift potential and the redshift are dimensionless.

As shown in Ref.\ \cite{Philipp:2017}, a definition of the relativistic geoid can be given in terms the redshift potential's level sets. 
These equipotential surfaces are called isochronometric, i.e.\ two standard clocks on the same level surface have a vanishing mutual redshift.
Therefore, the relativistic geoid is defined in Bjerhammar's philosophy as ``the surface on which all clocks run with the same speed'' but with more mathematical rigor and intrinsic general relativistic notions.
We want to point out here, that this definition remains valid in the following, but we will use new notions to make it accessible and also more intuitive with clear non-relativistic limits. 
Moreover, the definition will enable a direct comparison to existing post-Newtonian results and shed some light on the order of magnitude of deviations from the conventional geoid.

\subsection{Gravity Potential}
We assume that Einstein's field equation is fulfilled and the configuration is asymptotically flat and stationary.
Given a relativistic spacetime model of the Earth with a metric \eqref{Eq:metricStationary} that allows for the existence of a time-independent redshift potential $\phi$ for observers who are rigidly co-rotating, i.e.\ observers who form an isometric congruence, we construct the following.

\emph{Definition:} Let the relativistic gravity potential $U^*$ be defined by the following relation to the observers' time-independent redshift potential $\phi$ \cite{Philipp:2019}:
\begin{align}
	\label{Eq:UStarDefinition}
	e^{\phi} =: 1 + \dfrac{U^*}{c^2} 
	\,\, \Leftrightarrow \,\, 
	U^* = c^2 \left( e^{\phi} - 1 \right) = c^2 \left( \sqrt{-g_{00}} - 1 \right) \, ,
\end{align}
where we use co-rotating coordinates as defined above.
The dimension of $U^*$ is the square of a velocity, ${\left[U^*\right] = [c^2] = \text{m}^2 / \text{s}^2}$.
The intention of defining the new potential in this way becomes evident in the Newtonian limit, in which $U^*$ becomes the Newtonian gravity potential $W$, i.e.\
\begin{align}
	\label{Eq:UStarNewtonianLimit}
	U^* \underset{c \to \infty}{\longrightarrow} W \, .
\end{align}
This is easily verified by expanding the square root in Eq.\ \eqref{Eq:UStarDefinition} in the usual way, assuming a weak field limit exists.
Centrifugal effects are included since the coordinates are adapted to rigidly co-rotating observers such that they move on integral curves of the Killing vector field $\xi_{(1)}$.
Note in particular that in our sign convention also ${U^* < 0}$ everywhere. Hence, it has the same sign as Newton's gravity potential $W$.
A definition of the relativistic geoid in terms of $U^*$ will then also resemble the conventional Newtonian definition in terms of $W$ in the limit.

\subsection{Redshift and Acceleration}
Using the potential $U^*$, redshift and acceleration measurements for observers in the congruence can be expressed in the following way.
\subsubsection*{a) Redshift of two rigidly co-rotating observers}
Let the worldlines of two observers in the congruence be $\gamma_1$ and $\gamma_2$, respectively, and assume they measure their respective proper time, i.e.\ they are equipped with standard clocks. 
We evaluate the redshift potential $\phi$ and the relativistic gravity potential $U^*$ on their worldlines according to
\begin{align}
	\phi_i := \phi |_{\gamma_i} \quad \text{and} \quad U^*_i := U^*|_{\gamma_i} \quad i=1,2 \, .
\end{align}
The frequency ratio of the observers' clocks is then given by
\begin{align}
	1 + z &= \dfrac{\nu_1}{\nu_2} =  e^{\phi_2 - \phi_1} = \dfrac{1+U_2^*/c^2}{1+U_1^*/c^2} \notag \\
	&= 1 + \dfrac{U^*_2 - U^*_1}{c^2} + \mathcal{O}(c^4) =: 1 + \dfrac{\Delta U^*}{c^2} + \mathcal{O}(c^4) \, .
\end{align}
Hence, the relativistic gravity potential $U^*$ determines the redshift. 
To leading order, the redshift is given by the potential differences and vanishes in the Newtonian limit since Newton's universal time is absolute. 
But to first post-Newtonian order we obtain as the largest contribution
\begin{align}
	\dfrac{\nu_1}{\nu_2} = 1 + \dfrac{\Delta W}{c^2} + \mathcal{O} \big( c^{-3} \big) \, ,
\end{align}
since $U^*$ can be expressed as 
\begin{align}
	U^* = W + \sum_{n=2} \Xi_n/c^n \, ,
\end{align}
where the $\Xi_n$ are post-Newtonian correction terms of order $n$.

Note, however, that the definition of $U^*$ is exact, i.e. without any approximation, and it is valid for an arbitrary stationary spacetime. 
Thus, we have constructed an intrinsic general relativistic analog of the concepts introduced by Soffel \emph{et al.} in Ref.\ \cite{Soffel:1988}. 
In this work, the authors derive a similar potential but work within a first-order post-Newtonian approximation only.

\subsubsection*{b) Acceleration of freely falling objects w.r.t.\ rigidly co-rotating observers}
Using the fact that the acceleration potential of an isometric congruence is the same as its redshift potential, see Ref.\ \cite{Philipp:2017}, we can express the acceleration of freely falling test masses w.r.t.\ the observers in terms of $U^*$, 
\begin{align}
	a = -c^2 \mathrm{d}\phi = -c^2 \dfrac{\partial \phi}{\partial U^*} \mathrm{d}U^* = \dfrac{-\mathrm{d} U^*}{1+U^*/c^2} \, .
\end{align}
Here, $\mathrm{d}$ denotes the exterior derivative. The one form acceleration $a$ is closed and exact.
Its components can be calculated by ${a_\mu = -c^2 \partial_\mu \phi}$ and it is clear that ${a_0 \equiv 0}$. Thus, ${(a_\mu) = (0,a_1,a_2,a_3)}$ and the components are
\begin{align}
	a_i = -c^2 \partial_i \phi = -c^2 e^{-\phi} \partial_i e^{\phi} = \dfrac{- \partial_i U^* }{ 1 + U^*/c^2 } \, .
\end{align}
We notice that $U^*$, or rather its scaled gradient, determines the acceleration of freely falling test masses. The level surfaces $\phi = \text{const.}$, i.e.\ $U^* = \text{const.}$, are everywhere perpendicular to the acceleration -- that is perpendicular to the local plumb lines.

In the weak-field limit, we also recover the well-known Newtonian formula,\footnote{Note that in this limit the indices are raised and lowered with the Kronecker delta $\delta^\mu_\nu$.}
\begin{align}
	\vec{a} = - \vec{\nabla} W \, ,
\end{align}
according to which the gravity vector is determined by the gradient of the gravity potential.
For the magnitude of the relativistic acceleration we obtain
\begin{subequations}
\begin{align}
	\dfrac{a^2}{c^4} 
	&= \dfrac{g(a,a)}{c^4} = g^{ij} \partial_i \phi \, \partial_j \phi \\
	&\Rightarrow 
	a^2 = g^{ij} \dfrac{\partial_i U^* \, \partial_j U^*}{\left( 1+U^*/c^2 \right)^2} \, ,
\end{align}
\end{subequations}
from which the usual Newtonian definition of scalar gravity follows in the weak-field limit,\footnote{In all the equations above, $\vec{\nabla}$ is the flat space operator. Note that in geodesy, the magnitude of the acceleration -- gravity -- is usually denoted by $g$ \cite{TorgeMueller:2012}.}
\begin{align}
	a^2 = (\vec{\nabla} W)^2 \quad \Leftrightarrow \quad a = || \vec{\nabla} W ||  \, .
\end{align}
Here, $||\cdot||$ is the Euclidean norm.

\subsection{The Relativistic Geoid}
\emph{Definition:} For a spacetime equipped with a metric of the form 
\begin{align}
	g = \left(1 + \dfrac{U^*}{c^2} \right)^2 \left[ -(c \, \mathrm{d}t + \alpha_a(x) \mathrm{d}x^a)^2 + \alpha_{ab}(x) \mathrm{d}x^a \mathrm{d}x^b \right]
\end{align}
and a congruence of rigidly co-rotating observers who move on integral curves of the Killing vector field ${\xi_{(1)} = \partial_t}$, the relativistic geoid is a particularly chosen level surface of the relativistic gravity potential $U^*$ such that 
\begin{align}
\label{Eq:relGeoid}
U^* \big|_{\text{geoid}} = U^*_{0} = \text{const} \, .
\end{align}

Level surfaces of $U^*$ are level surfaces of $\phi$ and, therefore, they are isochronometric. 
The observers' worldlines are described by the tangent vector field ${u = (1+U^*/c^2)^{-1} \xi_{(1)}}$.
Consequently, in a stationary general relativistic model for the Earth's exterior, the relativistic geoid as determined by rigidly co-rotating observers on its surface by either clock comparison or plumb line directions is given by the particular two-dimensional isochronometric surface on which Eq.\ \eqref{Eq:relGeoid} holds.
The value of the redshift and gravity potential on a given isochronometric surface is invariant under coordinate transformations, respectively. The same is true for the redshift, which is related to potential differences.

The Newtonian limit of our general relativistic definition is apparent. Because the weak-field limit of $U^*$ is $W$, the Newtonian definition in terms of level surfaces of $W$ is recovered.
The value $U^*_{0}$, which singles out one equipotential surface, must be given by some convention. Two possibilities are, e.g., as follows:
\begin{itemize}
    \item[(i)] To fix the value by the conventional Newtonian gravity potential on the geoid such that ${U^*_{0} \equiv -W_0}$, 
    \item[(ii)] To define a master clock which is, by definition, situated on the geoid's surface and singles out one isochronometric surface in a geometrical way (e.g., the one that passes through its center of mass). Then, $U^*_0$ is related to the work done to bring a unit mass from infinity to the clock's position. 
\end{itemize}
Note that the choice (ii) is equivalent to marking a point (possibly at the shore), representing the mean sea level, and therefore choosing a level surface for $U^*$ in a geometric manner. 
Also, the construction of a so-called clock compass \cite{Puetzfeld:2018} might be employed to fix a particular isochronometric surface and to test all clocks w.r.t.\ it.

The definition of the geoid is exact and does not only apply to the Earth but also to arbitrarily compact objects as long as the requirements above are fulfilled. 
The geoid in various spacetimes can now be determined by expressing the respective metric in the form \eqref{Eq:metricStationary} and reading off the gravity potential.
In this way, also a comparison to previous results in a post-Newtonian framework is possible and presented in the next section.

\subsection{\label{Sec:Examples}Examples and Limits}
In this section, we show particular examples for the application of our definitions. 
With the spacetimes considered in the following, the influence of the relativistic monopole, quadrupole, and higher-order moments can be studied order by order. 
We start with a first-order parametrized post-Newtonian metric and proceed to exact Weyl solutions that include, in particular, the Schwarzschild and the Erez-Rosen spacetime. 
Moreover, results for the Kerr metric reveal the influence of gravitomagnetic contributions.
We argue that exact spacetimes play a useful role in relativistic geodesy and should be employed to explicitly define, understand, and calculate  conceptional notions and quantities, c.f.\ Ref.\ \cite{Soffel:2016}.

\subsubsection{Parametrized post-Newtonian Framework}
In harmonic co-rotating coordinates $(x^0 = c \, t, x,y,z)$, the metric components for the parametrized post-Newtonian spacetime that describes the Earth's exterior can be given by \footnote{We use only the parameters $\beta$ and $\gamma$ here.}
\begin{subequations}
\label{Eq:MetricPPN2}
\begin{align}
	g_{00}(\mathbf{x}) &= - \left( 1 + \dfrac{2 W(\mathbf{x})}{c^2} + \dfrac{2 \beta U(\mathbf{x})^2}{c^4} \right) + \mathcal{O}(c^{-6}) \, , \\
	g_{0i}(\mathbf{x}) &= - \, \dfrac{(\gamma+1) \, |U^i(\mathbf{x})|}{c^3}  - \dfrac{\epsilon_{ijk} {x}^j \omega^k}{c} + \mathcal{O}(c^{-5}) \, , \\
	g_{ij}(\mathbf{x}) &= \delta_{ij} \left( 1 - \dfrac{2\gamma U(\mathbf{x})}{c^2} \right) + \mathcal{O}(c^{-4}) \, ,
\end{align}
\end{subequations}
see, e.g., Refs.\ \cite{Soffel:1989, Kopeikin:Book:2011, Will:2014}.
Here, $\vec{\omega}$ is the Earth's rotation vector w.r.t.\ coordinate time, pointing in the $x^3$-direction, and $U^i$ is the post-Newtonian vector potential \cite{Will:2014, Poisson:2014}.
In these coordinates, rigidly co-rotating observers on the Earth's surface are described by ${\mathrm{d}x^i = 0}$.
The post-Newtonian approximation of General Relativity is obtained for $\beta =1$ and $\gamma = 1$.
Note that ${g_{0i}(\mathbf{x})}$ can also be expressed in the form
\begin{align}
	g_{0i}(\mathbf{x}) = - \dfrac{\gamma+1}{c^3} L_{\oplus,i} - \dfrac{\epsilon_{ijk} x^j \omega^k}{c} \, ,
\end{align}
with the Earth's gravitomagnetic field \cite{Soffel:1988, Will:1993}
\begin{align}
	\vec{L}_{\oplus} = \dfrac{G\vec{J}_\oplus \times \vec{x}}{R^3} 
\end{align}
and the total angular momentum ${\vec{J}_\oplus}$ of the Earth.
For a spherical central mass $M$ with radius $R_0$, rotating around the $x^3$-axis with angular velocity $\omega$, the gravitomagnetic vector potential can be evaluated easily in associated spherical coordinates, leading to
\begin{align}
	g_{t\Phi} = - \left( \dfrac{2(\gamma + 1)}{5} \left(\dfrac{R_0}{R}\right)^2 \dfrac{m}{R} + 1 \right)\omega R^2 \sin^2 \theta \, ,
\end{align}
which can also be derived by an expansion of the Kerr metric in the weak field for $\beta =1$.

With the definition \eqref{Eq:UStarDefinition}, the relativistic gravity potential $U^*_{\text{ppN}}$ for the parametrized post-Newtonian metric is 
\begin{align}
	\label{Eq:GravityPotentialPPN}
	U^*_{\text{ppN}} = W + \dfrac{U^2(\beta - 1/2)}{c^2} \, ,
\end{align}
and for the post-Newtonian approximation of General Relativity we get
\begin{align}
	\label{Eq:GravityPotentialPN}
	U^*_{\text{pN}} = W + \dfrac{U^2}{2c^2} \, .
\end{align}
Hence, deviations from the Newtonian gravity potential are described by the second term in Eq.\ \eqref{Eq:GravityPotentialPPN}, which is proportional to $U^2/c^2$.
Note that this result coincides with the findings in Refs.\ \cite{Soffel:1988, Kopeikin:2015}. 
However, the relativistic gravity potential $U^*$ here is defined without any approximations regarding the strength of the gravitational field. 
Thus, our framework also covers well-known results in an appropriately constructed limit of General Relativity.

Assume that signals are sent from one observer on the worldline $\gamma_1$ to another observers on the worldline $\gamma_2$, and both observers rigidly co-rotate with the Earth.
To the appropriate order ${\mathcal{O}(c^{-2})}$, the redshift is
\begin{align}
	\label{Eq:RedshiftPPN}
	1+ z 
	= \dfrac{\nu_1}{\nu_2} = \dfrac{1+U^*_2/c^2}{1+U^*_1/c^2}
	= 1 + \dfrac{W_2 - W_1}{c^2} + \mathcal{O}(c^{-4}) \, ,
\end{align}
where ${U^*_i := U^* |_{\gamma_i}}$ and ${W_i := W |_{\gamma_i}}$ for $i=1,2$.
To first order the redshift is proportional to $\Delta W := W_2 - W_1$ and it is not sensitive to the ppN parameters $\beta$ and $\gamma$. 
In fact, with Eq.\ \eqref{Eq:RedshiftPPN} we have just derived the fundamental  equation of chronometric geodesy.
Since we derived it in a top-down approach, we know how it is conceptually embedded in a broader theoretical framework; this gives trust in its validity and it is not only an approximative result which could, in principle, become useless at the full theoretical level.

The leading-order contribution to the redshift is due to the relativistic monopole moment and given by the result for the Schwarzschild spacetime; see below.
Close to the Earth's surface, a height difference of two clocks results in a redshift of roughly $10^{-16}$ per meter height distance.
For most applications related to clock comparison close to the Earth's surface, it will be sufficient to expand the gravity potential in Eq.\ \eqref{Eq:RedshiftPPN} up to quadrupolar order proportional to $J_2$ in Eq.\ \eqref{Eq:NewtonianPotentialExpansionAxisym}.
For a small spatial distance between the two clocks, the expansion of the gravity potential $W_2$ around the value $W_1$ leads to 
\begin{align}
	W_2 = W_1 + \vec{\nabla}W \cdot (\vec{X}_2 - \vec{X}_1) + \mathcal{O} \big( |\vec{X}_2 - \vec{X}_1|^2 \big) \, .
\end{align}
Hence, we obtain
\begin{align}
	W_2 = W_1 - \bar{g}_{12} H_{12} + \mathcal{O} \big( H_{12}^2 \big) \, ,
\end{align}
in terms of the orthometric height $H_{12}$ \cite{TorgeMueller:2012} between both clock positions. Here, $\bar{g}_{12} < 0$ denotes the averaged gravity between the clocks' positions along the plumb line.
Therefore, the redshift becomes
\begin{align}
	z\big( H_{12} \big) \approx |\bar{g}_{12}| \dfrac{H_{12}}{c^2}  \,\, > \, 0 \, .
\end{align}
Hence, redshift measurements are useful to determine the orthometric height, provided that they are supported by gravity measurements.
Indeed, geodetic leveling measurements must always be supported by matching gravity observations to conclude meaningful height measures \cite{TorgeMueller:2012, Philipp:2019}.

The relativistic geoid in the parametrized post-Newtonian spacetime is given by the equipotential surface of $U^*_{\text{ppN}}$ such that
\begin{align}
	U^*_{\text{ppN}} \big|_{\text{geoid}} = \left. W + \dfrac{U^2(\beta - 1/2)}{c^2} \right|_{\text{geoid}} = U^*_0 = \text{const.}
\end{align}
For the post-Newtonian approximation of General Relativity, we obtain 
\begin{align}
	U^*_{\text{pN}} \big|_{\text{geoid}} = \left. W + \dfrac{U^2}{2c^2} \right|_{\text{geoid}} = U^*_0 = \text{const.}
\end{align}
Thus, high-accuracy geoid determination might give bounds on the value of $\beta$.
The result above will be used later to access the leading-order relativistic corrections to the Newtonian geoid.
The results above are consistent with the literature; see Ref.\ \cite{Soffel:1988}.

We now calculate the covariant acceleration components $a_i$, the contravariant components $a^i$, and the norm of the acceleration $a$ at the level of $\mathcal{O}(1/c^2)$ accuracy,\footnote{Formally, the contravariant acceleration components can include a nonzero $a^0$. However, we have $a^0_{\text{ppN}} = 0 + \mathcal{O}(c^{-3})$.}
\begin{subequations}
\label{Eq:AccelerationPPN}
\begin{align}
	a_{i,\text{ppN}}
	&= - \partial_i \left( W + \dfrac{U^2 (\beta-1)}{c^2} \right) \, , \\
	a^i_{\text{ppN}}
	&= - \delta^{ij} \, \partial_j \left( W + \dfrac{U^2 (\beta + \gamma -1) }{c^2} \right) \, , \\ 
&\text{which is equivalent to }\notag \\
	\vec{a}_{\text{ppN}} 
	&= -\vec{\nabla} \left( W + \dfrac{U^2 (\beta + \gamma -1)}{c^2} \right) 
	=: -\vec{\nabla} \bar{U}_{\text{ppN}} \, \\
	a_{\text{ppN}}
	&= \left|\left| \vec{\nabla} \left( W + \dfrac{U^2 (\beta + \gamma/2 -1)}{c^2} \right) \right|\right| 
	=: || \vec{\nabla} \tilde{U}_{\text{ppN}} || \, .
\end{align}
\end{subequations}
Here, $\vec{\nabla}$ is the flat space operator such that ${|| \vec{\nabla} \tilde{U} || = \sqrt{\delta^{ij} \, \partial_i \tilde{U} \partial_j \tilde{U}}}$ in Cartesian coordinates. 

As shown in Eqns.\ \eqref{Eq:AccelerationPPN}, two new potentials are defined: (i) the potential $\bar{U}_{\text{ppN}}$, which determines the ``acceleration vector'' $\vec{a}_{\text{ppN}}$, and (ii) the potential $\tilde{U}_{\text{ppN}}$ of which the norm of the gradient gives the scalar acceleration $a_{\text{ppN}}$.
The results coincide with those in Ref.\ \cite{Soffel:1988}.

Note that for the parameter values of General Relativity $(\beta = 1, \gamma = 1)$, we have
\begin{subequations}
\begin{align}
	\bar{U}_{\text{pN}} &= W + \dfrac{U^2}{c^2} \, , \\
	\tilde{U}_{\text{pN}} &= U^*_{\text{pN}} = W + \dfrac{U^2}{2c^2} \, .
\end{align}
\end{subequations}
Interestingly, for the post-Newtonian approximation of General Relativity it is true that $\tilde{U}_{\text{pN}} \equiv U^*_{\text{pN}}$. 
However, measuring the acceleration of co-rotating observers on the Earth's surface yields bounds on the combination $(\beta + \gamma/2 -1)$.

In this section, we have shown how the results in Ref.\ \cite{Soffel:1988}, which was one of the major sources of inspiration for the development of our work, are included in the present formalism.
We regard it as an essential test of the framework that these results are successfully recovered - be reminded of our sign convention for the comparison.

\subsubsection{\label{Sec:Weyl} Weyl Metrics}
In the following, we apply our framework to Weyl metrics, which are axisymmetric and static solutions to Einstein's vacuum field equation.
We explicitly consider solutions which are asymptotically and elementary flat; see Ref.\ \cite{Quevedo:1989}.
Special examples of this class of spacetimes are the Schwarzschild solution and the quadrupolar spacetime found by Erez and Rosen \cite{Erez:1959}.
However, a straightforward generalization to higher-order multipole spacetimes exists.

Written in spheroidal non-rotating coordinates $(t,x,y,\varphi)$, the Weyl metric reads
\begin{multline}
	\label{Eq:WeylMetricSpheroidal}
	g_{\mu \nu} dx^{\mu} dx^{\nu}  = -e^{2\psi} c^2 dt^2 + m^2 e^{-2\psi} (x^2-1)(1-y^2)d\varphi^2
	\\
	+ m^2 e^{-2\psi} e^{2\gamma} (x^2-y^2) 
	\left( \dfrac{dx^2}{x^2-1} + \dfrac{dy^2}{1-y^2} \right) \, .
\end{multline}
The metric functions $\psi, \gamma$ depend only on $x$ and $y$, and $m$ is a length parameter determined by the total mass.
Einstein's vacuum field equation for the metric above can be found, e.g., in Refs.\ \cite{Quevedo:1989} and \cite{Stephani:Book:2003}.
In Ref.\ \cite{Quevedo:1989}, Quevedo has shown that the general asymptotically flat solution with elementary flatness on the axis is given by 
\begin{align}
	\label{Eq:WeylMetricExpansionSpheroidal}
	\psi = \sum_{l=0}^\infty (-1)^{l+1} q_l \, Q_l(x) \, P_l(y) \, .
\end{align}
Here, the $Q_l$ are Legendre functions of the second kind, see, e.g., Ref.\ \cite{Bateman:1955} for details.
We call the expansion coefficients $q_l$ Quevedo moments and they are related to Weyl's moments for the expansion in his canonical coordinates.
There is a clear relation to invariantly defined multipoles \footnote{We refer to the definition given by Geroch and Hansen \cite{Geroch:1970b, Hansen:1974}.} and to the Newtonian moments $J_l$ in the weak-field limit \cite{Philipp:2019},
\begin{align}
	J_l = (-1)^{l} \dfrac{l!}{(2l+1)!!} \left( \dfrac{m}{R_{\text{ref}}} \right)^l q_l \, ,
\end{align}
such that $m$ is the mass monopole in geometric units and $R_{\text{ref}}$ is a reference radius; see Eq.\ \eqref{Eq:NewtonianPotentialExpansionAxisym}.

\begin{figure*}
	\centering
	\includegraphics[width=1\textwidth]{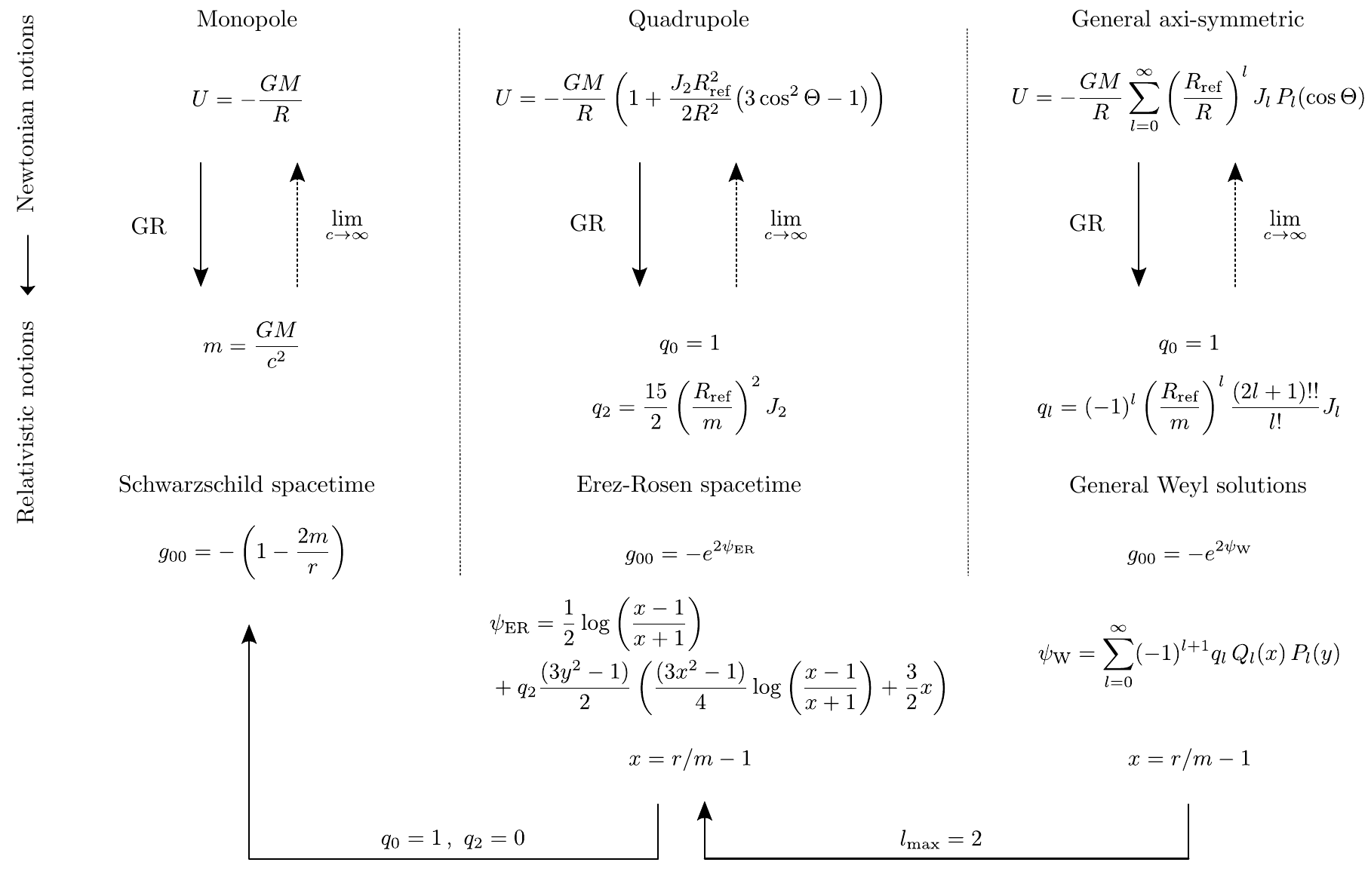}
	\caption{\label{Fig:Spacetimes}An overview of the relation between Weyl solutions and the Erez-Rosen as well as the Schwarzschild metric as generalizations of Newtonian configurations. An axisymmetric Newtonian gravitational field is generalized by a general Weyl solution with Quevedo moments $q_l$. For $l_\text{max} = 2$, the Erez-Rosen spacetime is recovered as the generalization of a Newtonian quadrupolar field. If also $q_2 = 0$, we obtain the Schwarzschild solution as the relativistic monopole analog.}
\end{figure*}

Changing to co-rotating coordinates allows us to include centrifugal effects. Then, we can read off the relativistic gravity potential from the final form of the metric.
This procedure includes inertial effects but of course fails to cover gravitomagnetic contributions since we still have a static spacetime at hand.
From the metric \eqref{Eq:WeylMetricSpheroidal} we conclude that two Killing vector fields $\xi_{(1)} = \partial_t$ and $\xi_{(2)} = \partial_\varphi$ exist, the latter of which is spacelike.
Rigidly co-rotating observers move on integral curves of ${\xi_{(1)} + \omega \xi_{(2)}}$, in which $\omega$ is the angular velocity of rotation around the symmetry axis. For bounded values of $\omega$ the combination remains timelike.
After the transformation $\varphi \to \varphi\prime = \varphi - \omega t$, we obtain the redshift potential for rigidly co-rotating observers,
\begin{align}
	\label{Eq:WeylRedshiftPotentialRot}
	e^{\phi(x,y)} = \sqrt{e^{2\psi(x,y)} - \dfrac{\omega^2}{c^2} m^2(x^2-1)(1-y^2) e^{-2\psi(x,y)}} \, .
\end{align}
Note how Weyl's first metric function $\psi$ enters the result.
Now, we can insert the exact result\footnote{The upper limit of summation $[l/2-1/2]$ denotes the closest integer smaller than the value of the expression in brackets.}
\begin{multline}
	\label{Eq:WeylRedshiftPotentialStat}
	2\psi(x,y) = \\
	\sum_{l=0}^n (-1)^{l+1} q_l P_l(y) 
	\times \left( \log \left( \dfrac{x+1}{x-1} \right) P_l(x) \right. \\
	\left. - 2 \sum_{k=0}^{[l/2-1/2]} \dfrac{2l-4k-1}{(l-k)(2k+1)} P_{l-2k-1}(x) \right) \, ,
\end{multline}
which allows us to evaluate the redshift potential at any multipolar level.
The very first term proportional to $q_0$ gives the Schwarzschild result, and $m$ is the Schwarzschild mass related to the radius $r_s = 2m$ for the choice $q_0 = 1$. To transform the result to usual area coordinates, the relation ${x= r/m-1,\, y = \cos \vartheta}$ must be used.
Including also the next higher-order term proportional to $q_2$ gives the redshift potential in the Erez-Rosen spacetime.\footnote{There is no relevant $q_1$ contribution since a dipole moment can always be made to vanish by a suitable coordinate transformation.}
Axisymmetric exact spacetimes with well-defined Newtonian limits can be constructed by including also $q_n$ for $n>2$; see Fig.\ \ref{Fig:Spacetimes} for a schematic overview.
The relativistic gravity potential for such Weyl spacetimes is a rather lengthy expression. 
However, it can be calculated without any approximation and is of the form
\begin{align}
	U^*_{\text{Weyl}} = U^*_{\text{Weyl}}(x,y,m,q_2 \dots q_l,\omega) \, .
\end{align}
The surfaces $U^* = \text{const.}$ are isochronometric and for a given set of parameters one of them is the geoid after a suitable choice of a constant $U^*_0$.
In the following, we give the explicit expressions for the Schwarzschild and Erez-Rosen spacetime, respectively.

The redshift potential for rigidly co-rotating observers in the Schwarzschild spacetime is
\begin{align}
	\label{Eq:SchwarzschildRedshiftPotentialRot}
	e^{\phi(r,\vartheta)} 
	&= \sqrt{-\big( g_{00}(r,\vartheta) + \omega^2 g_{\varphi\varphi}(r,\vartheta)/c^2 \big)} \notag \\
	&= \sqrt{1- \dfrac{2m}{r} - \dfrac{\omega^2}{c^2} r^2 \sin^2\vartheta} \, ,
\end{align}
where we use standard area coordinates.
The relativistic gravity potential for the Schwarzschild spacetime becomes
\begin{align}
	U^*_{\text{Schwarzschild}} = c^2 \left( \sqrt{1- \dfrac{2GM}{c^2r} - \dfrac{\omega^2}{c^2} r^2 \sin^2\vartheta} - 1 \right) \, .
\end{align}

Hence, the redshift between two members of the congruence at positions ${(r_1,\vartheta_1)}$ and ${(r_2,\vartheta_2)}$, respectively, is
\begin{align}
	\label{Eq:SchwarzchildZ}
	z+1 
	= \dfrac{\nu_1}{\nu_2}
	= \dfrac{\sqrt{1- \dfrac{2m}{r_2} - \dfrac{\omega^2}{c^2} r_2^2 \sin^2\vartheta_2}}{\sqrt{1- \dfrac{2m}{r_1} - \dfrac{\omega^2}{c^2} r_1^2 \sin^2\vartheta_1}} \, ,
\end{align} 
and we find that close to the Earth's surface the redshift is about $10^{-18}$ per cm height difference.
For the magnitude of the acceleration along the congruence, we find
\begin{multline}
	a = e^{-2\phi} \left( \left( 1-\dfrac{2GM}{c^2r} \right) \left( \dfrac{GM}{r^2} - \omega^2 r \sin^2\vartheta \right)^2 \right. \\
	\left. + \dfrac{1}{r^2} \left(\omega^2 r^2 \sin \vartheta \cos \vartheta \right)^2 \right)^{1/2} \, .
\end{multline}
In the Newtonian limit, this becomes
\begin{align}
	a = ||\vec{\nabla} W || \, , \quad \text{with} \quad W = -\dfrac{GM}{R} - \dfrac{1}{2} \omega^2 R^2 \sin^2 \Theta \, ,
\end{align}
which is the gravity magnitude in a spherically symmetric field.

The Erez-Rosen spacetime can be used to describe the spacetime outside a quadrupolar Earth. 
It is a natural relativistic generalization of a quadrupolar Newtonian gravitational potential,
\begin{align}
	\label{Eq:NewtonianPotentialQuadrupole}
	U(R,\Theta) 
	= - \dfrac{GM}{R} \left( 1 + \dfrac{J_2 R_{\text{ref}}^2}{2R^2} \big( 3\cos^2 \Theta - 1 \big) \right) \, ,
	\end{align}
which will be recovered in the weak-field limit.
We can choose the parameters $m$ and $q_2$ to let the relativistic monopole $\mathcal{M}_0$ and the quadrupole $\mathcal{M}_2$ coincide with the respective Newtonian moments of the Earth.
To do so, we have to choose ${m_\oplus = GM/c^2}$ and $q_2$ must be
\begin{align}
	q_2 = \dfrac{15}{2} \left( \dfrac{R_{\text{ref}}}{m} \right)^2 J_2 \, .
\end{align}
The resulting redshift potential reads
\begin{multline}
	\label{Eq:Erez-RosenRedshiftPotentialRot}
	e^{\phi(x,y)} = \\
	\sqrt{ e^{2\psi_{\text{ER}}(x,y)} - \dfrac{\omega^2}{c^2} (GM/c)^2(x^2-1) (1-y^2) e^{-2\psi_{\text{ER}(x,y)}} } \, ,
\end{multline}
in which the metric function $\psi_{\text{ER}}$ is given by 
\begin{multline}
	\psi_{\text{ER}(x,y)} = \dfrac{1}{2}\log \left( \dfrac{x-1}{x+1} \right) 
	+ q_2 \dfrac{(3y^2-1)}{2} \\
	\times \left( \dfrac{(3x^2-1)}{4} \log \left( \dfrac{x-1}{x+1} \right) + \dfrac{3}{2} x \right) \, .
\end{multline}
Using the expressions above, the relativistic gravity potential $U^*_{\text{ER}}(x,y,m,q_2,\omega)$ for the Erez-Rosen spacetime can be calculated analytically and the geoid is defined by a level surface in some chosen convention for $U^*_0$.

\subsubsection{Kerr Spacetime}
To investigate the influence of gravitomagnetic contributions, we apply the framework to the Kerr spacetime.
In the standard Boyer-Lindquist coordinates, the metric is given by
\begin{align}
	\label{Eq:KerrMetricBL}
	g = 
	&- \left( 1-\dfrac{2mr}{\rho^2} \right) c^2 \mathrm{d}t^2 
	+ \dfrac{\rho^2}{\Delta} \mathrm{d}r^2 
	+ \rho^2 \mathrm{d}\vartheta^2 \notag \\
	&+ \sin^2 \vartheta \left( r^2 + a^2 + \dfrac{2m r a^2\sin^2\vartheta}{\rho^2} \right) \mathrm{d}\varphi^2 \notag \\ 
	&- \dfrac{4mra\sin^2\vartheta}{\rho^2} \, c \, \mathrm{d}t \mathrm{d}\varphi \, , 
\end{align}
and we introduce the abbreviations
\begin{align}
	\rho^2 = r^2 + a^2 \cos^2 \vartheta \, , \quad \Delta = r^2 + a^2 -2mr \, .
\end{align}
The gravitomagnetic field of the Kerr spacetime approximates frame dragging effects in the Earth's vicinity very well, if parameters are chosen appropriately.
The first multipole moments of the Kerr metric are
\begin{subequations}
\begin{align}
	\text{Mass monopole}& \quad &&\mathcal{M}_0 = m \, , \\
	\text{Spin dipole}& \quad &&\mathcal{S}_1 = ma \, , \\
	\text{Mass quadrupole}& \quad &&\mathcal{M}_2 = -m a^2 \, .
\end{align}
\end{subequations}
The Kerr parameter is related to the angular momentum by $a = J/(Mc)$.
Hence, we can choose $m$ such that the Kerr monopole is the total mass of the Earth, and $a$ such that the spin dipole is related to the Earth's angular momentum.
To have this correspondence, we set ${m_\oplus = GM/c^2}$, as before, and use the relation of the angular momentum to the moment of inertia $I$, ${J = I \omega}$.
For a rigidly rotating sphere with radius $r_\oplus$, we have $I = 2/5 M r_\oplus^2$.
Hence, the Kerr parameter for the Earth becomes
\begin{align}
	a_\oplus = \dfrac{2}{5} \dfrac{\omega}{c} r_\oplus^2  \, .
\end{align}
Calculating the values, we obtain $m_\oplus \approx \SI{0.0044}{\meter}$ and ${a_\oplus \approx 892 \, m_\oplus}$, where the radius and angular velocity as given by the EGM96 model are used \cite{Lemoine:1998}\footnote{Note that for this value the Kerr spacetime is actually over-extreme. However, this is true for any planet or star and horizons / singularities are of no interest since they are in the interior but the solution is valid and used for the exterior part only.}.
However, we have to state that the Kerr spacetime is also not a good approximation for the Earth's exterior in the sense that it covers gravitomagnetic effects but fails to represent the Earth's flattening and quadrupole moment.
Choosing the values for the mass monopole and the spin-dipole fixes the mass quadrupole uniquely. 
Therefore, we cannot expect the Kerr spacetime to represent features of the flattened Earth.

We change again to rotating coordinates to find the relativistic gravity potential for rigidly co-rotating observers,
\begin{align}
	\label{Eq:KerrRedshiftPotentialRot}
	U^*_{\text{Kerr}} / c^2 = 
	&-1 + \left( 1 - \dfrac{2mr}{\rho(r,\vartheta)^2} 
	+ 4 \, \dfrac{\omega}{c} \,  \dfrac{amr\sin^2\vartheta}{\rho(r,\vartheta)^2} \right. \notag \\
	&\left. - \, \dfrac{\omega^2}{c^2} \, \sin^2 \vartheta \left( r^2+a^2+\dfrac{2m r a^2\sin^2\vartheta}{\rho(r,\vartheta)^2} \right) \right)^{1/2} \, .
\end{align}
This potential can be used to compute redshifts, accelerations and the relativistic geoid in the Kerr spacetime.
For $a \to 0$ the Schwarzschild result is recovered.
Gravito-magnetic effects are included in two terms. One is proportional to $a \omega$ and the other is proportional to $a^2 \omega^2$.
The first term can change the sign depending on the direction of rotation, whereas the latter mixes with inertial effects.

\section{Magnitudes of Relativistic Effects on the Geoid}
In this section, we determine and quantify the magnitudes of relativistic corrections to the Earth's geoid at the leading order. 
We study a simple quadrupolar Earth model in the Newtonian theory and its relativistic generalization in the framework presented above.
The best relativistic generalization of such a model is given by an appropriately constructed Erez-Rosen spacetime; see Sec.\ \ref{Sec:Weyl}.
However, leading-order effects are given by the first-order post-Newtonian approximation of the Erez-Rosen metric, which is the post-Newtonian metric constructed for a quadrupolar Newtonian potential \eqref{Eq:NewtonianPotentialQuadrupole}. 
This spacetime, constructed with the help of Eqns.\ \eqref{Eq:MetricPPN2}, will be employed in the following.

The Earth's quadrupole moment\footnote{Note that both are positive in our convention, whereas in geodesy, a different sign convention applies and $C_{20} \equiv - J_2$} $C_{20} \equiv J_2$, related to its flattening, gives the first (and by far the largest) nontrivial contribution to its gravitational field beyond the monopole.
The quadrupole causes gravity to change from the equator toward the poles with a $\sin^2 \alpha$-like behavior, where $\alpha$ is the geocentric latitude.
Hence, we expect relativistic corrections to (i) induce an overall spherical correction due to the monopole, and (ii) yield latitude-dependent corrections about three orders of magnitude smaller since ${J_2/J_0 \approx 10^{-3}}$ for the Earth.

The methodology of this section is as follows. 
We use the Newtonian quadrupolar gravity potential \eqref{Eq:NewtonianPotentialQuadrupole} and, thereupon, construct the post-Newtonian approximation of this situation to access the first-order relativistic contributions. 
Then, we construct the Newtonian geoid, based on the gravity potential $W$ and the relativistic geoid based on $U^*$. 
In either case, we obtain a two-dimensional surface given by some function $x^1(x^2)$, where $x^1$ is a radial coordinate and $x^2$ is related to the polar angle.

The comparison of both results must be done in a way that eliminates coordinate ambiguities. 
We have decided to use an isometric embedding of the relativistic geoid surface into the three-dimensional Euclidean space $\mathbb{R}^3$. 
If such an embedding is possible, it is unique and allows us to investigate the intrinsic geometry of the relativistic geoid by applying well-known methods for the analysis of curved two-dimensional surfaces.
The Newtonian geoid generically ``lives'' in this Euclidean space and a comparison of two-dimensional surfaces in $\mathbb{R}^3$ is possible, e.g., in terms of their radial distance in any angular direction.
Therefore, once the relativistic geoid is embedded, we can compare it to the Newtonian one and determine the difference.
Note that such an embedding is in general only possible by numerical methods, even though the embedding equations can be given analytically; see Appendix for details.

We also have to overcome some subtle conventional issues.
In the Newtonian case, the geoid is defined by the level surface of the gravity potential such that ${|W| = W_0}$ on its surface. 
Nowadays, $W_0$ is an agreed upon constant related to coordinate time transformations from TCG to TAI scales; see Ref.\ \cite{Sanchez:2016}.
Already in the Newtonian case, there are conceptional difficulties with properties of the geoid being ``a mean sea surface fit'' and the derived constant $W_0$ being not at all directly related to the sea surface.
Let us, therefore, assume that some value $W_0$ is chosen, one way or another, that defines the Newtonian geoid.
In the relativistic case, we define the geoid by one particular isochronometric surface such that ${U^* |_{\text{geoid}} = U^*_{0}}$. 
Now, we need a clear prescription of how to choose the value $U^*_0$ and how to relate it to $W_0$. 
Hence, some gauge freedom is left in the choice of the constant. 
In the following, we consider different approaches and calculate the differences between the Newtonian and relativistic geoid in either case.
 
In approach (I), we choose $-U^*_0 \equiv W_0$, which may be obvious regarding the Newtonian limit and is supported by the results of the previous section, as well as the proper time on the geoid and the defining constant $L_g$ in the IAU resolution with its relation to $W_0$.
 
In approach (II), we choose the value of $U^*$ such that after the isometric embedding into $\mathbb{R}^3$ and comparison to the Newtonian geoid, the difference vanishes in the equatorial plane and is globally as small as possible. 
This means, we use the gauge freedom in the comparison such that the relativistic geoid is as close as possible to the Newtonian geoid.

In approach (III), we choose the value of $U^*_0$ such that in its post-Newtonian expansion $W \to W_0$ and $U \to U_0$.

Finally, we also consider approach (IV), which is analog to approach (I) but without embedding the relativistic geoid into $\mathbb{R}^3$. 
Instead, we identify the global coordinates that are used for the Newtonian geoid and the relativistic post-Newtonian metric.
This approach allows to judge whether or not the embedding is really necessary at the leading order or merely an academic endeavor.

%-----------------------------------
%	SUBSECTION
%-----------------------------------

\subsection{Geoid models}
The Newtonian gravity potential of a quadrupolar gravitational field is
\begin{align}
	W(R,\Theta) = - G \left( \dfrac{M}{R} + \dfrac{ N_2 P_2 (\cos \Theta) }{R^3} \right) - \dfrac{1}{2} R^2 \Omega^2 \sin^2 \Theta  \, ,
\end{align}
where ${N_2 = J_2 R_\oplus^2 M_\oplus}$. 
We use spherical coordinates $(R,\Theta,\Phi)$ for the $\mathbb{R}^3$, and $R_\oplus$, $M_\oplus$ are the Earth's radius and mass, respectively, and $\Omega$ its angular velocity. 
The geoid as determined by observers on the rotating Earth, i.e.\ including centrifugal effects, in this model is the level surface of $W$ such that
\begin{align}
	-W(R,\Theta) |_{\text{geoid}} = W_0 = \SI{6.26368534e7}{\square\meter \per \square\second}\, .
\end{align}

For the post-Newtonian approximation of this configuration, we have to use
\begin{align}
	U^*(r,\theta) 	
	&= U(r,\theta) - \dfrac{1}{2} r^2 \Omega^2 \sin^2\theta + \dfrac{1}{2} \dfrac{U(r,\theta)^2}{c^2} \notag \\
 	&= W(r,\theta) + \dfrac{1}{2} \dfrac{U(r,\theta)^2}{c^2} \, .
\end{align}
The relativistic geoid is given by one chosen level surface of $U^* (r,\theta)$ such that
\begin{align}
	U^* (r,\theta) |_{\text{geoid}} = U^*_0 \, .
\end{align}
We choose pseudo-spherical coordinates ${(r,\theta,\varphi)}$ for the post-Newtonian spacetime and the flat space coordinates ${(R,\Theta,\Phi)}$ are their 0th order approximations.

%-----------------------------------
%	SUBSECTION
%-----------------------------------

\subsection{Geoid comparison approach (I)}

Using the comparison method (I), we choose 
\begin{align}
	U^*_0 = - W_0 
\end{align}
and construct the relativistic geoid for the post-Newtonian configuration. 
The result is a two-surface described by $r(\theta)$.
After embedding this surface into $\mathbb{R}^3$, we determine the radial distance, at any polar angle, to the Newtonian geoid, which is generically given in $\mathbb{R}^3$ as a function $R(\Theta)$. 
The result is shown in Fig.\ \ref{Fig:GeoidDiffPN-N}. 
We find the mean difference between both geoids to be about $2\,$mm with some small angular deviations that are three orders of magnitude smaller due to the quadrupolar influence. 
Hence, we exactly find what we predicted at the beginning: the relativistic monopole causes a global deviation and the relativistic quadrupole induces some angular variation.

In a nutshell, the first-order corrections to the Earth's geoid due to General Relativity are about $2\,$mm with latitudinal variations of about $3\,\mu\text{m}$.

%-----------------------------------
%	SUBSECTION
%-----------------------------------

\subsection{Geoid comparison approach (II)}
For the second approach, we choose the value $U^*_0$ such that in the embedding space both geoids coincide in the equatorial plane.
This choice can be easily translated into the structure of the embedding equations; see appendix.
However, in general it is not possible to deduce properties or parameter values in the spacetime from requirements which shall be fulfilled after an embedding.
First, the equatorial radius $R_0$ of the Newtonian geoid is calculated and, thereupon, the solution $r_0(R_0)$ of $g_{\varphi\varphi}(r_0,\pi/2) = R_0^2$ is used as initial condition for the embedding equations; see Eq.\ \eqref{Eq:Embedding_PN_b}.
Thus, it is guaranteed that in the equatorial plane the embedded relativistic geoid has the same radius as the Newtonian one.
In terms of the constant $U^*_0$, this relates to the choice 
\begin{align}
	U^*_0 &= -W_0 - \SI{0.02}{\square\meter \per \square\second}  \notag \\
	&= -\SI{6.26368534e7}{\square\meter \per \square\second} - \SI{0.02}{\square\meter \per \square\second} \, .
\end{align}
Hence, there is a small difference between $U^*_0$ and $W_0$ in this gauge, which is indeed below the contemporary accuracy of measurements for $W_0$ but corresponds to the next significant digit.

\begin{figure*}[tb]
	\centering
	\includegraphics[width=\textwidth]{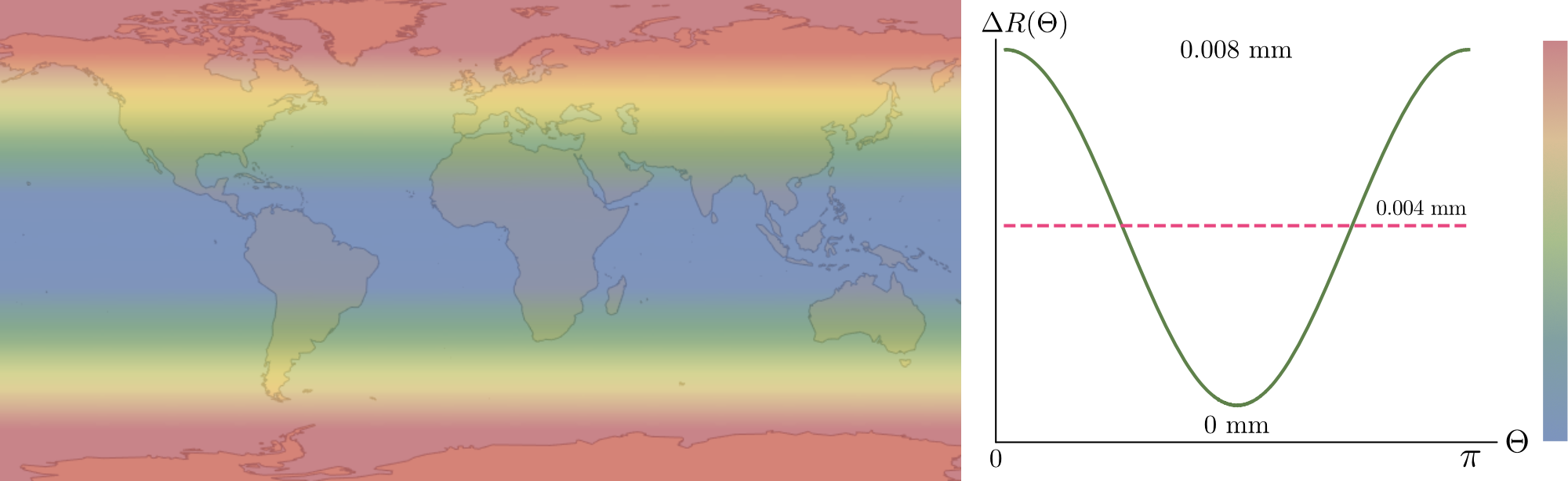}
	\caption[Comparison of the relativistic and Newtonian geoids: approach (II).]{\label{Fig:GeoidDiffPN-N_b}Comparison of the relativistic and Newtonian geoid at leading order for approach (II). We show the geoid radii differences $\Delta R(\Theta) = R_{\text{PN}}(\Theta) - R_{\text{N}}(\Theta)$ in the embedding space $\mathbb{R}^3$ as a function of $\Theta$. The maximal, minimal, and mean differences are indicated.}
\end{figure*}

The result of this approach is shown in Fig.\ \ref{Fig:GeoidDiffPN-N_b}.
We see that the overall modulation is almost completely removed, the mean value is about $4\,\mu\text{m}$, and only the latitudinal variation of about $8\,\mu\text{m}$ remains.
By this choice, we have fitted the relativistic geoid to the Newtonian one in the best possible way and the influence of the relativistic monopole was reduced (or corrected for) as good as possible.
The remaining difference is mainly due to the relativistic quadrupole.
Hence, the freedom in the choice of gauge and convention can be used to minimize the differences between both geoids.
Note, however, that for this approach it is no longer true that $U^*_0 = -W_0$ and the numerical values that define the surfaces of the relativistic and conventional geoid, respectively, are different.
No obvious \emph{a priori} choice for the constant $U^*_0$ can be made and this limits the geometric realization by measurement stations.

%-----------------------------------
%	SUBSECTION
%-----------------------------------

\subsection{Geoid comparison approach (III)}
\begin{figure*}[tb]
	\centering
	\includegraphics[width=\textwidth]{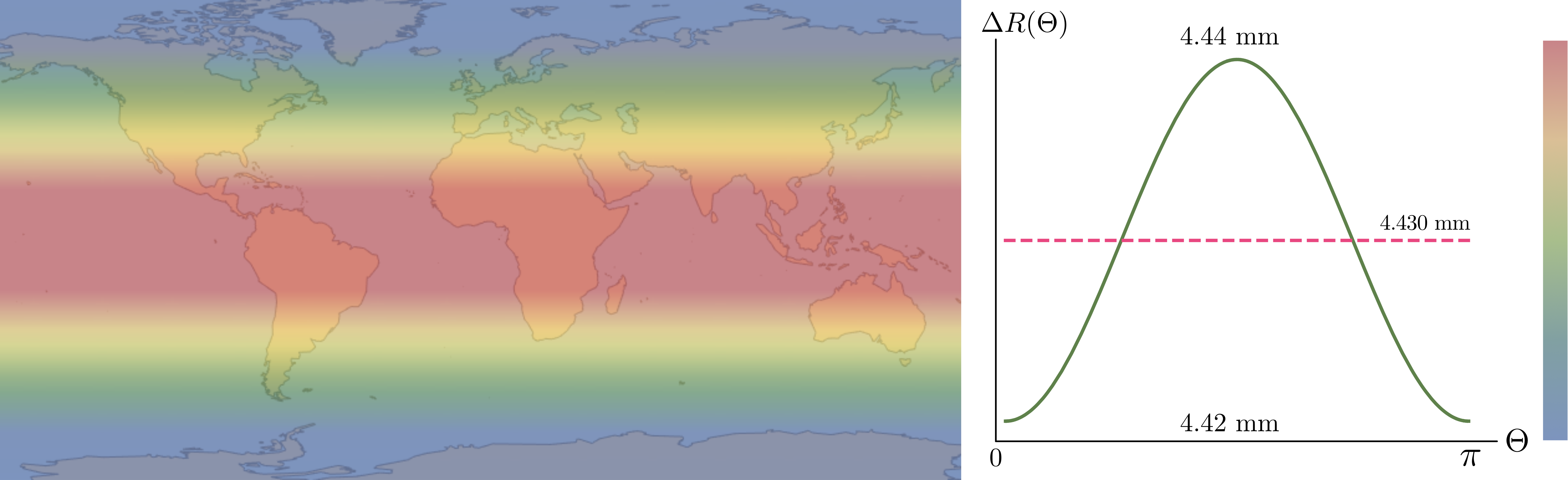}
	\caption[Comparison of the relativistic and Newtonian geoids: approach (III).]{\label{Fig:GeoidDiffPN-N_c}Comparison of the relativistic and Newtonian geoid at leading order for approach (III). We show the geoid radii differences $\Delta R(\Theta) = R_{\text{PN}}(\Theta) - R_{\text{N}}(\Theta)$ in the embedding space $\mathbb{R}^3$ as a function of $\Theta$. The maximal, minimal, and mean differences are indicated.}
\end{figure*}

For the third approach, $U^*_0$ is chosen such that 
\begin{align}
	 U^*_0 = U^* \big|_{\text{geoid}} = U^*_\text{pN} \big|_{W \to -W_0, U \to U_0} = -W_0 + \dfrac{U_0^2}{2c^2} \, ,
\end{align}
and the result is shown in Fig.\ \ref{Fig:GeoidDiffPN-N_c}.
For this choice, the difference between Newtonian and relativistic geoid in the embedding space is about $4\,$mm with latitudinal variation of $0.02\,$mm between the poles and the equatorial plane.
The disadvantage of this choice is that the value of $U^*_0$ varies for each post-Newtonian order. 
Hence, we may exclude choices like the this from further analysis.

%-----------------------------------
%	SUBSECTION
%-----------------------------------

\subsection{Geoid comparison approach (IV)}

\begin{figure*}[tb]
	\centering
	\includegraphics[width=\textwidth]{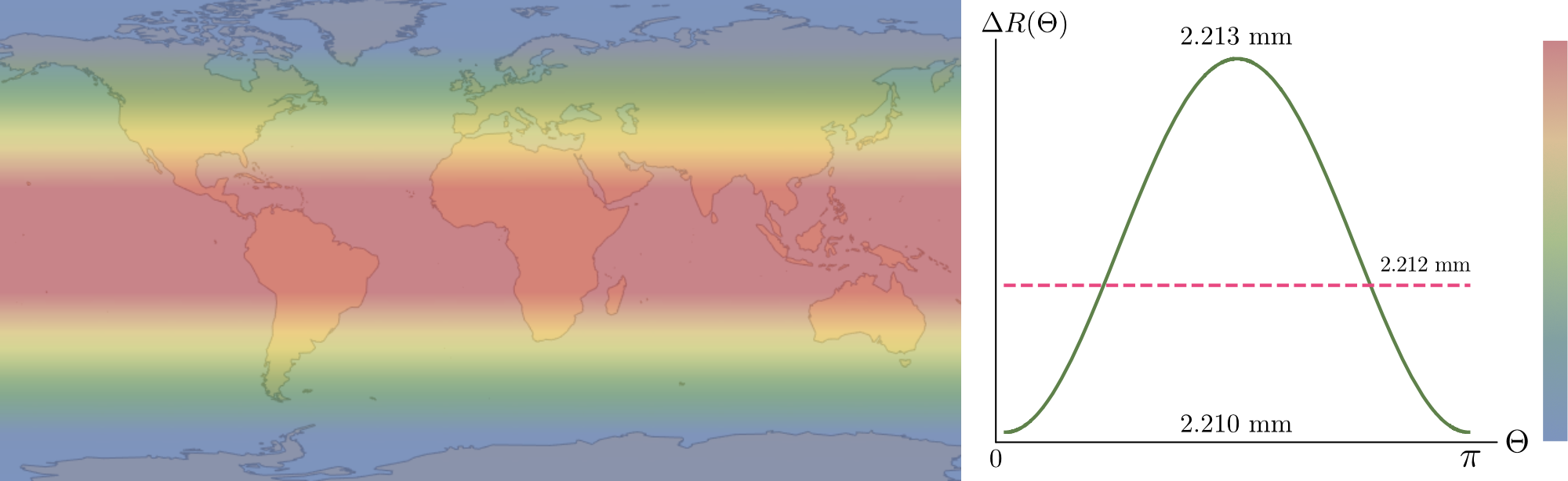}
	\caption[Comparison of the relativistic and Newtonian geoids: approach (I).]{\label{Fig:GeoidDiffPN-N}Comparison of the relativistic and Newtonian geoid at leading order for approach (I). We show the geoid radii differences $\Delta R(\Theta) = R_{\text{PN}}(\Theta) - R_{\text{N}}(\Theta)$ in the embedding space $\mathbb{R}^3$ as a function of $\Theta$. The maximal, minimal, and mean differences are indicated.}
\end{figure*}

\begin{figure*}[tb]
	\centering
	\includegraphics[width=\textwidth]{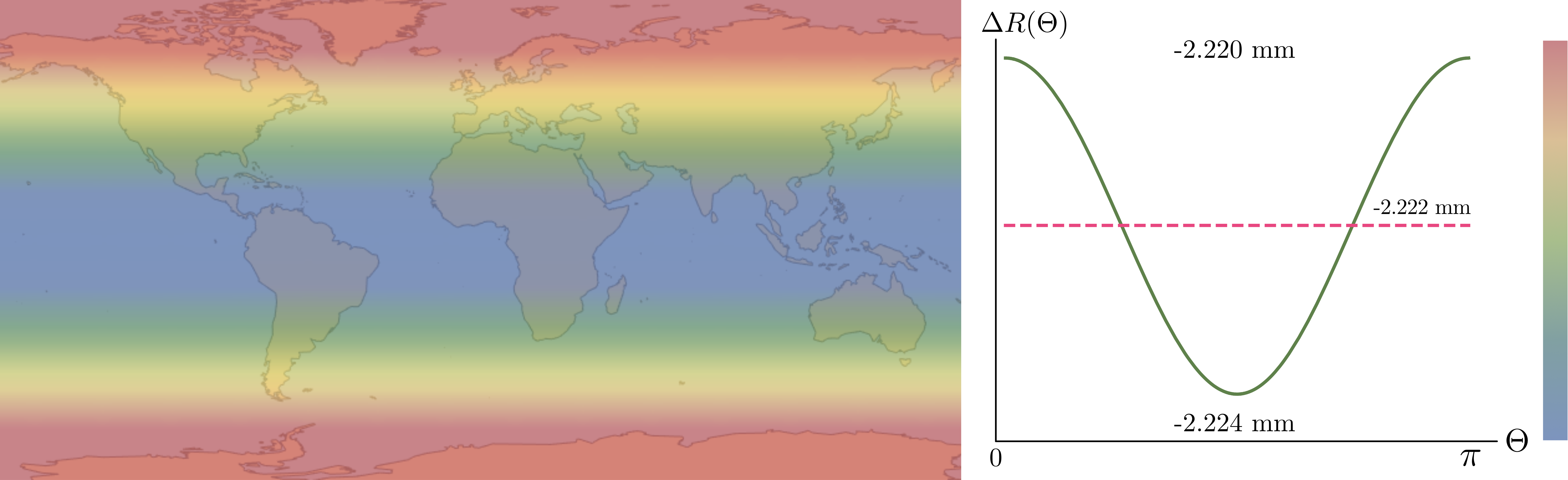}
	\caption[Comparison of the relativistic and Newtonian geoids: approach (IV).]{\label{Fig:GeoidDiffPN-N-NoEmbedding}Comparison of the relativistic and Newtonian geoid at leading order for approach (IV), which is the same as approach (I) but without embedding. The Newtonian and post-Newtonian coordinates are identified. We show the geoid radii differences $\Delta R(\Theta) = R_{\text{PN}}(\Theta) - R_{\text{N}}(\Theta)$ as a function of $\Theta$. The maximal, minimal, and mean differences are indicated.}
\end{figure*}

To emphasize the importance of the embedding, we compare the relativistic and Newtonian geoid also without any embedding but for identification of the coordinates. 
The result is quite remarkable and shown in Fig.\ \ref{Fig:GeoidDiffPN-N-NoEmbedding}.
We clearly see that, w.r.t.\ approach (I), the sign of the difference changes.
As we can infer from Fig.\ \ref{Fig:GeoidDiffPN-N} for approach (I), the radius of the relativistic geoid is globally about $2\,$mm larger than the radius of the Newtonian geoid.
For approach (IV) it is vice versa!
The ``effect'' appears due to the mismatch of the coordinates which is of the order of $4\,$mm at the involved distances,  such that pN radii are 'smaller' than Newtonian radii.
Hence, the embedding is not only a theorists pedantism but really has an important influence on the result.
It is, however, a mere coincidence that $+2\,$mm becomes $-2\,$mm by coordinate and embedding effects. 
Therefore, statements on the magnitude of ``$2\,$mm difference'' between the geoids, as communicated in the geodetic community, remain correct; but signs do matter.


\section{Conclusion and Outlook}
We have shown how the definition of a relativistic gravity potential leads to a definition of the relativistic geoid in analogy to the Newtonian understanding and expands the results that were presented in terms of the time-independent redshift potential in Ref.\ \cite{Philipp:2017}.
Our framework contains previously published results and generalizes Newtonian and post-Newtonian notions developed by other authors.
The leading-order difference of $2\,$mm between relativistic and conventional geoids might be detected with clocks and redshift measurements at the $10^{-19}$ level in the future. 
Therefore, clock networks using transportable optical clocks as well as clock stations and fiber networks can establish a realization of the relativistic geoid; see Refs. \cite{Droste:2013, Lisdat:2016, Takano:2016, Mueller:2017, Guena:2017, Delva:2017b, Grotti:2018, Mehlstaeubler:2018}.
Moreover, our geoid definition can also be realized by acceleration measurements and the determination of local plumb lines.
For the realization of reference systems, the Global Geodetic Observing System (GGOS) of the International Association of Geodesy (IAG) demands $1\,$mm as accuracy and $0.1\,$ mm/yr for its stability \cite{Plag:2009}. 
Thus, even today the relativistic geoid should be adopted in practice to fulfill these requirements and to be consistent with the relativistic treatment of the other geodetic methods (space geodetic techniques, reference frames, Earth rotation, etc.).

In follow-up publications we will focus on incorporating gravitomagnetic effects by expanding our results to spacetimes that allow to resemble the Earth's monopole, quadrupole, and spin dipole at the same time.
Therefore, we need to overcome the limitation of the Kerr metric, in which the mass quadrupole is uniquely fixed by the choice of the spin dipole, and use another class of spacetime models allowing for the choice of more free parameters.
Moreover, building on the present framework, we will generalize other geodetic notions such as the normal gravity field, height definitions for chronometric geodesy and relativistic leveling, and timescales as well as their relation to the proper time on the geoid.

\section*{Acknowledgement}
We gratefully acknowledge the financial support by the Deutsche Forschungsgemeinschaft (DFG) under Germany’s Excellence Strategy EXC-2123/1 (Project-ID: 390837967) and through the Research Training Group 1620 ``Models of Gravity''.
This work was also supported by the German Space Agency DLR with funds provided by the Federal Ministry of Economics and Technology (BMWi) under Grant No.\ DLR 50WM1547 and the DLR institute for Satellite Geodesy and Inertial Sensing.
The authors would like to thank Volker Perlick, Dirk Puetzfeld, Heiner Denker, Domenico Giulini, and Sergei Kopeikin for helpful discussions.
\bibliography{geoid.bib}
\appendix
\section{Isometric Embedding into Euclidean Space}
To investigate and visualize the intrinsic geometry of the relativistic geoid, which is described by a particular isochronometric surface, i.e.\ a level surface of the relativistic gravity potential $U^* |_{\text{geoid}} = U^*_0 = \text{const.}$, we isometrically embed this surface into Euclidean space $\mathbb{R}^3$. If such an embedding is possible, the embedded surface shows the intrinsic geometry of the geoid.

In all our axisymmetric models, the relativistic geoid is a level surface,
\begin{align}
	\label{Eq:Appendix:Embedding1}
  	U^*(x,y) |_{\text{geoid}} = U^*_0 = \text{const.} \, ,
\end{align}
where $x$ and $y$ are the two spatial coordinates, related to a radius measure and the polar angle, respectively.

For any such two-dimensional surface defined by Eq.\ \eqref{Eq:Appendix:Embedding1}, the following is true everywhere on the surface 
\begin{align}
	\label{Eq:appendix:Embedding2}
  	0 = \mathrm{d}U^* =  \partial_x U^*(x,y) \, \mathrm{d}x + \partial_y U^*(x,y) \, \mathrm{d}y \, .
\end{align}
Hence, we have on the geoid surface
\begin{align}
	\label{Eq:appendix:Embedding3}
  	\mathrm{d}x = -\left( \dfrac{\partial_y U^*(x,y)}{\partial_x U^*(x,y)} \right) \mathrm{d}y \, ,
\end{align}
which yields a relation $x=x(y)$ that describes this surface.

On the surface $U^*(x,y) = U^*_0$ there is a two-dimensional Riemannian metric defined according to
\begin{align}
	\label{Eq:appendix:Embedding4}
	g^{(2)} = \left( g_{xx}(x,y) \, x'(y)^2 + g_{yy}(x,y) \right) \mathrm{d}y^2 + g_{\varphi \varphi}(x,y) \, \mathrm{d}\varphi^2 \, ,
\end{align} 
where $\varphi$ is the azimuthal angle related to the axisymmetry of the spacetime model.

We want to isometrically embed the surface $U^*(x,y) = U^*_0$ into Euclidean three-space $\mathbb{R}^3$ with cylindrical coordinates $(\Lambda,\varphi,Z)$, where $Z$ is the height, $\Lambda$ is the radius in the $Z=0$ plane, and $\varphi$ is the azimuthal angle. The Riemannian metric in $\mathbb{R}^3$ can then be written as
\begin{align}
	\label{Eq:appendix:Embedding5}
  	g_E^{(3)} = dZ^2 + d\Lambda^2 + \Lambda^2 d\varphi^2 \, .
\end{align}
The two embedding functions $Z(y)$ and $\Lambda(y)$ can now be determined from 
\begin{align}
	\label{Eq:appendix:Embedding6}
  	&\left[ g_{xx}(x(y),y) \, x'(y)^2 + g_{yy}(x(y),y) \right] \mathrm{d}y^2 + g_{\varphi \varphi}(x(y),y) \, \mathrm{d}\varphi^2 \notag \\
  	&= \left( Z'(y)^2 + \Lambda'(y)^2 \right) \, dy^2 + \Lambda^2(y) d\varphi^2 \, .
\end{align}
If Eq.\ \eqref{Eq:appendix:Embedding3} allows an explicit solution for $x=x(y)$, the result is inserted into \eqref{Eq:appendix:Embedding6}. A comparison of coefficients leads to
\begin{subequations}
\begin{align}
	\label{Eq_Lambda}
  	\Lambda(y) 	
  	&= \left. \sqrt{g_{\varphi \varphi}(x,y)} \right|_{x=x(y)} \, ,\\
         \label{Eq_h}
  	Z(y) 
  	&= \pm \int_{0}^{y} dy \, \left( g_{xx}(x,y) \left( \dfrac{\partial_y U^*(x,y)}{\partial_x U^*(x,y)} \right)^2 + g_{yy}(x,y) \right. \notag \\
  	&\left. -  \dfrac{g'_{\varphi \varphi}(x,y)^2}{4 g_{\varphi \varphi}(x,y)} \right)^{1/2}_{x=x(y)} \, .
\end{align}
\end{subequations}
Here, $g'_{\varphi \varphi}$ means that $x(y)$ is inserted first and then the derivative w.r.t.\ $y$ is taken. 
In general, Eq.\ \eqref{Eq_h} can only be integrated numerically.
 
The solutions of Eqs.\ \eqref{Eq_Lambda} and \eqref{Eq_h} yield the cylindrical radius coordinate $\Lambda$ and the height $Z$ in $\mathbb{R}^3$ as functions of $y \in [-1,1]$. 
This corresponds to a polar angle $\vartheta \in [0,\pi]$. 
Using the height and radius functions, a section of the embedded surface is obtained. Due to the axisymmetry, the full embedded surface is obtained by a rotation of this section.
For all values of $y$ for which it is true that
\begin{align}
  	g_{xx}(x,y) \left( \dfrac{\partial_y U^*(x,y)}{\partial_x U^*(x,y)} \right)^2 + g_{yy}(x,y)  > \dfrac{g'_{\varphi \varphi}(x,y)^2}{4 g_{\varphi \varphi}(x,y)} \, ,
\end{align}
the embedding is possible.
If the embedding into $\mathbb{R}^3$ is not possible, different means of visualization and comparison of the relativistic geoid must be employed.

If Eq.\ \eqref{Eq:appendix:Embedding3} cannot be solved for $x=x(y)$, we have to use Eq.\ \eqref{Eq:appendix:Embedding2}, which leads to
\begin{align}
  	x'(y) = \dfrac{dx}{dy} = - \dfrac{\partial_y U^*(x,y)}{\partial_x U^*(x,y)} \, .
\end{align}
Together with Eqs.\ \eqref{Eq_Lambda} and \eqref{Eq_h}, the following coupled system of differential equations must be solved
\begin{subequations}
	\begin{align}
  		x'(y) 
  		&= - \dfrac{\partial_y U^*(x,y)}{\partial_x U^*(x,y)} \, , \label{Eq:Embedding_PN_b} \\
  		\Lambda(y) 
  		&= \sqrt{g_{\varphi\varphi}(x(y),y)} \, , \\
  		Z'(y) &= \pm \sqrt{ g_{xx}(x(y),y) \left( x'(y) \right)^2 + g_{yy}(x(y),y) - \Lambda'(y)^2 } \, .
\end{align}
\end{subequations}
The initial conditions can be given in the equatorial plane $(y=0)$ such that ${ x(0) = x_0}$, ${Z(0) =0}$. 

\end{document}